\documentstyle[here,epsfig]{mn2e}

\def\gtsima{$\; \buildrel > \over \sim \;$}
\def\ltsima{$\; \buildrel < \over \sim \;$}
\def\gtrsim{\lower.5ex\hbox{\gtsima}}
\def\lesssim{\lower.5ex\hbox{\ltsima}}

\begin{document}

\title[Low metallicity and ULXs]{Low metallicity and ultra-luminous X-ray sources in the Cartwheel galaxy}
\author[Mapelli, Colpi, Zampieri]
{M. Mapelli$^{1}$, M. Colpi$^{2}$, L. Zampieri$^3$
\\
$^{1}$ Institute for Theoretical Physics, University of Z\"urich, Winterthurerstrasse 190, CH-8057, Z\"urich, Switzerland; {\tt mapelli@physik.unizh.ch}\\
$^2$Universit\`a Milano Bicocca, Dipartimento di Fisica G.Occhialini, Piazza
delle Scienze 3, I-20126, Milano, Italy\\
$^3$INAF-Osservatorio astronomico di Padova, Vicolo dell'Osservatorio 5, I-35122, Padova, Italy\\
}
\maketitle \vspace {7cm }

  \begin{abstract}
Low-metallicity ($Z\lesssim{}0.05\,{}Z_\odot{}$) massive ($\gtrsim{}40\,{}M_\odot{}$) stars might end their life by directly collapsing into massive black holes (BHs, $30\lesssim{}m_{\rm BH}/M_\odot{}\lesssim{}80$). More than $\sim{}10^5$ massive BHs might have been generated via this mechanism in the metal-poor ring galaxy Cartwheel, during the last $\sim{}10^7$ yr. We show that such BHs might power most of the ultra-luminous X-ray sources (ULXs) observed in the Cartwheel. We also consider a sample of ULX-rich galaxies and we find a possible anti-correlation between the number of ULXs per galaxy and the metallicity in these galaxies. However, the data are not sufficient to draw any robust conclusions about this anti-correlation, and further studies are required.
\end{abstract}

%
%________________________________________________________________

\section{Introduction}
Ultra-luminous X-ray sources (ULXs, see Mushotzky 2004 for a review, and references therein) are defined as point-like sources with isotropic X-ray luminosity $L_{\rm X}\gtrsim{}10^{39}$ erg s$^{-1}$, that is higher than the Eddington luminosity for a $\sim{}7\,{}M_\odot{}$ black hole (BH). Most of the brightest ULXs are located in starburst galaxies (Irwin, Bregman \&{} Athey  2004).
%found in many nearby spiral, irregular and elliptical galaxies (see Mushotzky 2004
%for a review, and references therein), but most of them (and especially the brightest) reside in starburst galaxies (Irwin, Bregman \&{} Athey  2004). 
The origin of ULXs is still an open question. Many different scenarios have been proposed. ULXs could be associated with high-mass X-ray binaries (HMXBs) powered by stellar-mass BHs with anisotropic X-ray emission (e.g. King et al. 2001) 
%(e.g. King et al. 2001; King \&{} Pounds 2003) 
or with super-Eddington accretion rate/luminosity (e.g. Begelman 2002) or with a combination of the two mechanisms (e.g. King 2008).
%(King \&{} Pounds 2003; Socrates \&{} Davis 2006; Poutanen et al. 2007). 
ULXs could also be associated with HMXBs powered by intermediate-mass BHs (IMBHs), i.e. BHs with mass $100\,{}M_\odot{}\le{}m_{\rm BH}\le{}10^5\,{}M_\odot{}$ (see van der Marel 2004 for a review). 
%However, for most of ULXs it is not necessary to invoke exotic scenarios, such as IMBHs. 
However, IMBH masses larger than $100\,{}M_\odot{}$ 
are not needed to explain the observational properties of most of ULXs (e.g. Gon\c{c}alves \&{} Soria 2006). IMBHs may be required only to explain the properties of some peculiar ULXs, such as the brightest (i.e. the $\lesssim{}4$ ULXs with $L_{\rm X}\gtrsim{}10^{41}$ erg s$^{-1}$), those which show quasi-periodic oscillations (M82 X-1, see Strohmayer \&{} Mushotzky 2003, and NGC 5408 X-1, see Strohmayer et al. 2007) or which are surrounded by isotropically ionized nebulae (e.g. Kaaret, Ward \&{} Zezas 2004). 

The Cartwheel galaxy, which is a starburst (Marston \& Appleton 1995; Mayya et al. 2005) and a collisional ring galaxy (Struck-Marcell \&{} Higdon 1993; Struck et al. 1996; Mapelli et al. 2008a, 2008b), hosts a particularly large population of ULXs ($\sim{}17$, Gao et al. 2003; Wolter \&{} Trinchieri 2004; Wolter, Trinchieri \&{} Colpi 2006). 
Recent studies (King 2004; Mapelli et al. 2008a) suggest that IMBHs can hardly account for all the ULXs observed in the Cartwheel. In fact, more than $\sim{}1000$ IMBHs are required in order to produce the 17 observed ULXs. Such high number of IMBHs is hard to produce according to the most common theoretical models, such as the runaway collapse in young stellar clusters (Portegies Zwart \&{} McMillan 2002), the repeated mergers of stellar-mass BHs in star clusters (Miller \&{} Hamilton 2002) or the remnants of population III stars (Heger et al. 2003, hereafter H03). In this Letter we investigate an alternative  scenario for the formation of massive BHs ($30\,{}M_\odot{}\le{}m_{\rm BH}\le{}80\,{}M_\odot{}$), which  could account for most of the ULXs in the Cartwheel and in other metal-poor starburst galaxies. 
This model is based on the idea that low-metallicity ($Z\sim{}0.05\,{}Z_\odot{}$, i.e. approximately the metallicity of the Cartwheel, Fosbury \& Hawarden 1977) massive stars ($\gtrsim{}40\,{}M_\odot{}$) lose only a small fraction of their mass due to stellar winds (Maeder 1992, hereafter M92; H03) and can directly collapse (Fryer 1999) into massive BHs ($30\,{}M_\odot{}\le{}m_{\rm BH}\le{}80\,{}M_\odot{}$).  This scenario has already been suggested in previous studies (Pakull \&{} Mirioni 2002; Zampieri et al. 2004; Soria et al. 2005; Swartz, Soria \&{} Tennant 2008), which pointed out a correlation between formation of ULXs and low-metallicity environments, and proposed that this may be connected with the influence of metallicity on the evolution of massive stars. The hypothesis that bright ULXs contain $\approx 30-90
M_\odot$ BHs formed in a low-metallicity environment 
%and accreting in a slightly critical regime 
is also considered in a
companion investigation (Zampieri \&{} Roberts 2009).
%ULX and Cartwheel ULX
%problem of formation

\section{Model}
According to numerical models (Fryer 1999; H03), a star which, at the end of its life, has a mass of $\ge{}40\,{}M_\odot{}$ is likely to directly collapse into a BH\footnote{However, depending on the metallicity, stars which undergo a strong luminous-blue-variable phase end their life as neutron stars (see Belczynski, Kalogera \& Bulik 2002 and references therein).}. In this case, the mass of the remnant BH is likely close to the final mass of the progenitor star, as no
significant mass ejection is expected in the direct collapse.
%progenitor, as no significant mass losses are expected in the direct collapse.
Massive stars with metallicity close to solar cannot have masses larger than $\sim{}10-15\,{}M_\odot{}$ at the end of their life, even if their initial mass was very large, as they are expected to lose a lot of mass due to stellar winds (H03; Meynet \&{} Maeder 2003). Instead, massive stars with lower metallicity are less affected by stellar winds and retain a larger fraction of their initial mass. M92 shows that a star with  metallicity $Z\sim{}0.05\,{}Z_\odot{}$ retains $\sim{}100$ per cent of its initial mass $m_{\rm in}$ if $m_{\rm in}\le{}40\,{}M_\odot{}$, and $\sim{}80-67$ per cent of $m_{\rm in}$ if $60\,{}M_\odot{}\le{}m_{\rm in}\le{}120\,{}M_\odot$. 
Thus, combining the results by M92 and by Fryer (1999), stars with metallicity $Z\lesssim{}0.05\,{}Z_\odot{}$ and initial mass $40\,{}M_\odot{}\le{}m_{\rm in}\le{}120\,{}M_\odot{}$ might end their life directly collapsing into BHs with mass $30\,{}M_\odot{}\le{}m_{\rm BH}\le{}80\,{}M_\odot{}$. Such BHs (that, in the following, we will dub simply as 'massive BHs') may be considered IMBHs, although close to the low-mass limit for IMBHs, and are sufficiently massive to power most of ULXs. 
%%can be hardly reconciled with 
%%are not needed to explain the observational properties of most of ULXs (Goncalves \&{} Soria 2006; Stobbart, Roberts \&{} Wilms 2006; Copperwheat et al. 2007; Roberts 2007). Thus, IMBHs with mass $30\,{}M_\odot{}\le{}m_{\rm BH}\le{}80\,{}M_\odot{}$ match reasonably well the properties of most ULXs.
%On the other hand, IMBH masses larger than $100\,{}M_\odot{}$  are not needed to explain the observational properties of most of ULXs (Goncalves \&{} Soria 2006; Stobbart, Roberts \&{} Wilms 2006; Copperwheat et al. 2007). Thus, massive BHs with mass $30\,{}M_\odot{}\le{}m_{\rm BH}\le{}80\,{}M_\odot{}$ match reasonably well the properties of most ULXs.
If this model is correct, we can approximately estimate the total number of massive BHs  (${\rm N}_{\rm BH}$) which are formed by this process during a burst of star formation (SF), as
\begin{equation}\label{eq:totnum}
{\rm N}_{\rm BH}=A\,{}\,{}\,{}\int_{40\,{}M_\odot{}}^{m_{\rm max}}m^{-\alpha{}}\,{}{\rm d}m,
\end{equation}
where $m_{\rm max}$ is the maximum stellar mass (we assume $m_{\rm max}=120\,{}M_\odot{}$) and  $\alpha{}$ is the index of the initial mass function (IMF). $A$, the normalization constant, can be estimated as
\begin{equation}\label{eq:norm}
A=\frac{{\rm SFR}\,{}\,{}\,{}t_{\rm burst}}{\int_{m_{\rm min}}^{m_{\rm max}}m^{1-\alpha{}}\,{}{\rm d}m}, 
\end{equation}
where SFR is the SF rate during the burst, $t_{\rm burst}$ the duration of the burst and $m_{\rm min}$ the minimum stellar mass (we assume $m_{\rm min}=0.08$ $M_\odot{}$).
%, $m_{\rm max}$ the maximum stellar mass (we assume 120 $M_\odot{}$) and $\alpha{}$ the index of the initial mass function (IMF). From equation~(\ref{eq:totmass}) 
Similarly, we can estimate the total mass of massive BHs (${\rm M}_{\rm BH}$) as
\begin{equation}\label{eq:totmass}
{\rm M}_{\rm BH}=A\,{}\,{}\,{}\int_{40\,{}M_\odot{}}^{m_{\rm max}}m^{-\alpha{}}\,{}\left(m\,{}b+c\right)\,{}{\rm d}m,
\end{equation}
where $b=0.54$ and $c=15.59\,{}M_\odot{}$ account for the mass losses due to stellar winds and have been derived by linearly fitting the values in table~1 of M92 for initial stellar masses $m_{in}\ge{}40\,{}M_\odot{}$. 
%Once ${\rm N}_{\rm BH}$ is known, one can estimate the efficiency of massive BHs, $\epsilon{}_{\rm BH}$, defined as the ratio between the number of observed ULXs in a galaxy (${\rm N}_{\rm ULX}$), assuming that all of them are powered by a massive BH, and the number of massive BHs inferred from our model (${\rm N}_{\rm BH}$):
Once ${\rm N}_{\rm BH}$ is known, we can estimate the upper limit ($\epsilon{}_{\rm BH}$) of the fraction of massive BHs which power ULXs in a given galaxy at present, assuming that all the observed ULXs in this galaxy are powered by a massive BH: 
\begin{equation}\label{eq:eobs}
\epsilon{}_{\rm BH}=\frac{{\rm N}_{\rm ULX}}{{\rm N}_{\rm BH}}.
\end{equation}
In order to check the robustness of this model,  $\epsilon{}_{\rm BH}$ can be compared with the fraction of massive BHs which are expected to power ULXs at present ($\epsilon{}_{\rm exp}$), derived combining recent dynamical (Blecha et al. 2006, hereafter B06) and binary-evolution (Patruno et al. 2005) models. Such models are completely independent and unrelated to the scenario presented in this Letter. In particular, B06 show that a massive BH with mass $\approx{}100\,{}M_\odot{}$ hosted in a young stellar cluster undergoes mass transfer from a companion star for a fraction $f_{\rm MT}\sim{}0.03$ of the life of the cluster\footnote{The value of $f_{\rm MT}$ derived from B06 accounts at statistical level for all the companions which may undergo mass transfer (i.e. both main sequence and post-main sequence).}. Patruno et al. (2005) show that only mass transfer between a massive BH and a star with mass $\ge{}10\,{}M_\odot{}$ is able to produce a persistent ULX, whereas, if the companion mass is lower ($2\,{}M_\odot{}\le{}m<10\,{}M_\odot{}$), the X-ray source is transient, with a very short burst (few days) every few months (Portegies Zwart, Dewi \&{} Maccarone 2004). A transient source reaches ULX luminosities only during the burst phase (Portegies Zwart et al. 2004). Thus, $\epsilon{}_{\rm exp}$ can be derived, on the basis to these models, as:
\begin{equation}\begin{array}{l}\label{eq:eexp}
\epsilon{}_{\rm exp}=f_{\rm MT}\,{}\left(\int_{m_{\rm min}}^{m_{\rm max}}m^{-\alpha{}}\,{}{\rm d}m\right)^{-1}\,{}\times{}\\\nonumber
\hspace{1.2cm}\left(\int_{10\,{}M_\odot{}}^{m_{\rm max}}m^{-\alpha{}}\,{}{\rm d}m\,{}+\,{}f_{\rm duty}\,{}\int_{2\,{}M_\odot{}}^{10\,{}M_\odot{}}m^{-\alpha{}}\,{}{\rm d}m\right),
\end{array}\end{equation}
where $m_{\rm min}$ and $m_{\rm max}$ are the same as adopted in eq.~(\ref{eq:norm}). $f_{\rm duty}$ represents the fraction of time which a transient source spends in its burst phase. In the following, we will assume $f_{\rm duty}=10^{-2}$, which is a reasonable upper limit (Portegies Zwart et al. 2004; King 2004). If $\epsilon{}_{\rm BH}$ is close  to $\epsilon{}_{\rm exp}$, this would indicate that our model provides reasonable results. We stress that this is a simple estimate of $\epsilon{}_{\rm exp}$, and has various limitations. First, the assumption that $f_{\rm duty}$ is constant is a simplification. On the other hand, there are still large uncertainties about the duty cycle and its dependence on the properties of the accreting system (Portegies Zwart et al. 2004). As we will show in the next Section, a different duty cycle does not affect significantly our estimates. Another limit of this model is the assumption that the massive BHs remain inside their parent cluster. Observations show that there is a displacement between some ULXs and the star clusters (e.g. Zezas et al. 2002). This might indicate that some massive BHs were ejected, together with their companion stars, from the parent clusters. In case of ejection, $f_{\rm MT}$ likely depends only on the evolution of the companion star (which cannot be exchanged with other stars) and is probably different from the estimates by B06. A new model would be required to quantify $f_{\rm MT}$ in case of ejection. This issue will be addressed in a forthcoming paper.
%Furthermore, the percentage of ULXs which are outside clusters needs to be estimated. Both these issues will be addressed in a forthcoming paper. 
However, the possible ejection of some BHs from the parent cluster does not affect the number of massive BHs which form in our scenario, as derived in eq.~(\ref{eq:totnum}), and the corresponding value of $\epsilon{}_{\rm BH}$ in eq.~(\ref{eq:eobs}), but only the estimate of $\epsilon{}_{\rm exp}$ in eq.~(\ref{eq:eexp}).
%%%{\bf We also note that the analytical estimates reported here may be affected by variations of metallicity across galaxies and multiple star formation episodes.}

%, enable us to derive an expected  efficiency $\epsilon{}_{\rm exp}$ In order to check the robustness of this model, the observed  efficiency $\epsilon{}_{\rm obs}$ can be compared with the expected  efficiency $\epsilon{}_{\rm exp}$, derived combining other dynamical and stellar evolution models.

\section{Results for the Cartwheel}
The Cartwheel, which has a low metallicity ($Z\sim{}0.05\,{}Z_\odot{}$, measured in the nebulae of the outer ring which are forming stars right now, Fosbury \&{} Hawarden 1977) and hosts a large number of ULXs ($\sim{}17$, Wolter \& Trinchieri 2004), appears the ideal candidate to check this model. First, let us estimate the approximate total number and mass of massive BHs which can form in the Cartwheel via such a mechanism, by using eqs.~(\ref{eq:totnum})$-$(\ref{eq:totmass}). The SFR in the Cartwheel is $\sim{}20\,{}M_\odot{}$ yr$^{-1}$ (Mayya et al. 2005). The time of the burst $t_{\rm burst}$ is probably the most uncertain among the quantities in eq.~(\ref{eq:norm}). Simulations show that $\sim{}100$ Myr have elapsed from the galaxy interaction which produced the Cartwheel's ring (Mapelli et al. 2008a). Thus, one can take  $t_{\rm burst}=10^8$ yr as an upper limit. However, we are not interested in all the massive BHs, but only in those which can easily acquire a massive stellar companion, that is those which are still in the parent stellar cluster. Furthermore, in order to produce the persistent ULXs which have been detected in the Cartwheel (Wolter \&{} Trinchieri 2004; Wolter et al. 2006), the parent star cluster should still host sufficiently massive stars. Thus, we also adopt
%, as a reference value, 
$t_{\rm burst}=10^7$ yr, which is approximately the lifetime of a 15 $M_\odot$ star. For the IMF in eq.~(\ref{eq:totnum}) we consider two different cases: a Salpeter IMF ($\alpha{}=2.35$, Salpeter 1955) and a Kroupa IMF, which is relatively top-heavy ($\alpha{}=1.3$ if $m\le0.5\,{}M_\odot{}$ and $\alpha{}=2.3$ for larger masses, Kroupa 2001).

The results from eqs.~(\ref{eq:totnum}) and (\ref{eq:totmass}) are the following (see Table~1). Assuming $t_{\rm burst}=10^8$ yr, the total number of massive BHs is N$_{\rm BH}=1.2\times{}10^6$ and N$_{\rm BH}=2.4\times{}10^6$ for the Salpeter and the Kroupa IMF, respectively. The total number of massive BHs born during the last $10^7$ yr is N$_{\rm BH}=1.2\times{}10^5$ and N$_{\rm BH}=2.4\times{}10^5$ for the Salpeter and the Kroupa IMF, respectively. These numbers are quite higher than those predicted  by the runaway collapse (Portegies Zwart \&{} McMillan 2002). In fact, even assuming that each massive ($\gtrsim{}10^4\,{}M_\odot{}$) young cluster produces one or even two IMBHs (G\"urkan, Fregeau \&{} Rasio 2006) via runaway collapse (which is an upper limit, see Gvaramadze, Gualandris \&{} Portegies Zwart 2008), $\sim{}10^5$ massive young clusters should form during the starburst, in order to generate the same number of IMBHs. Our model does not suffer from such limitations, as it predicts that more than one massive BH may form in the same cluster and that massive BHs can form also outside clusters. For a Salpeter (Kroupa) IMF, 
%the total mass of massive BHs which have formed during  $t_{\rm burst}=10^8$ yr is M$_{\rm BH}\sim{}6.2\times{}10^7\,{}M_\odot{}$ (M$_{\rm BH}\sim{}1.23\times{}10^8\,{}M_\odot{}$), whereas 
the total mass of massive BHs which are born during the last $10^7$ yr, and thus are able to produce a ULX, is  M$_{\rm BH}\sim{}6.2\times{}10^6\,{}M_\odot{}$ (M$_{\rm BH}\sim{}1.23\times{}10^7\,{}M_\odot{}$). The average mass of a single massive BH is $\langle{}m_{\rm BH}\rangle{}=50.2\,{}M_\odot{}$ and $\langle{}m_{\rm BH}\rangle{}=50.4\,{}M_\odot{}$, using the Salpeter and the Kroupa IMF, respectively.

Since the observed ULXs in the Cartwheel are 17, from eq.~(\ref{eq:eobs}) we obtain $\epsilon{}_{\rm BH}=1.4\times{}10^{-4}$ and $\epsilon{}_{\rm BH}=6.9\times{}10^{-5}$ for the Salpeter and the Kroupa IMF, respectively, if $t_{\rm burst}=10^7$ yr is assumed.
Let us see now how $\epsilon{}_{\rm BH}$ compares with $\epsilon{}_{\rm exp}$. 
From eq.~(\ref{eq:eexp}) we get $\epsilon{}_{\rm exp}=4.6\times{}10^{-5}$ and $\epsilon{}_{\rm exp}=2.4\times{}10^{-4}$ for a Salpeter and a Kroupa IMF, respectively. We stress that most of the contribution in  eq.~(\ref{eq:eexp}) comes from the persistent ULXs (those with a companion mass larger than $10\,{}M_\odot{}$). In fact, the value of $\epsilon{}_{\rm exp}$ once we neglect the transient sources is $\epsilon{}_{\rm exp}=4.3\times{}10^{-5}$ and $\epsilon{}_{\rm exp}=2.3\times{}10^{-4}$ for a Salpeter and a Kroupa IMF, respectively. Thus, a lower value of $f_{\rm duty}$ does not affect our results.
%%%%%%%%%%%%%%%%%%%%%%%%%%%%%%% TABLE 1%%%%%%%%%%%%%%%%%%%%%%%%%%%%%%%%%
\begin{table}
\begin{center}
\caption{Results for the Cartwheel (assuming $t_{\rm burst}=10^7$~yr).} \leavevmode
\begin{tabular}[!h]{lll}
\hline
& Salpeter 
& Kroupa \\
\hline
N$_{\rm BH}$              & $1.2\times{}10^5$ & $2.4\times{}10^5$ \\

M$_{\rm BH}$  ($M_\odot{}$)      &      $6.2\times{}10^6$ & $1.23\times{}10^7$\\
$\epsilon{}_{\rm BH}$        &  $1.4\times{}10^{-4}$ & $6.9\times{}10^{-5}$\\
$\epsilon{}_{\rm exp}$ &$4.6\times{}10^{-5}$& $2.4\times{}10^{-4}$\\
\noalign{\vspace{0.1cm}}
\hline
\end{tabular}
\end{center}
%\footnotesize{ $^{\rm a}$Assuming $t_{\rm burst}=10^7$ yr.
%}
\end{table}
%%%%%%%%%%%%%%%%%%%%%%%%%%%%%%%%%%%%%%%%%%%%%%%%%%%%%%%%%%%%%%%%%%%%%%%%%%%%%
The main conclusion is that $\epsilon{}_{\rm BH}$ and $\epsilon{}_{\rm exp}$ 
%the two estimates obtained with the two different and independent methods 
are quite similar one to the other. In particular, $\epsilon{}_{\rm BH}$ is $\sim{}3$ times higher than $\epsilon{}_{\rm exp}$ assuming a Salpeter IMF, and $\sim{}3$ lower than $\epsilon{}_{\rm exp}$ assuming a Kroupa IMF  (see Table~1). Thus, the model of massive BH formation from direct collapse of low-metallicity massive stars may be able to explain the ULXs observed in the Cartwheel.

\section{Comparison with other galaxies}
%%%%%%%%%%%%%%%%%%%%%%%%%%%%%%% TABLE 2%%%%%%%%%%%%%%%%%%%%%%%%%%%%%%%%%
\begin{table*}
\begin{center}
\caption{Properties of the galaxies in our sample.} \leavevmode
\begin{tabular}[!h]{lllll}
\hline
Galaxy
& SFR ($M_\odot{}$ yr$^{-1}$)
& $Z$ ($Z_\odot{}$)$^{\rm a}$
& N$_{\rm ULX}$
& references$^{\rm b}$\\
\hline
Cartwheel          & 20         & 0.05 (outer ring spectra)       & 17  & 1, 2, 3\\
AM~0644$-$741       & 3          & $0.45$ (bulge spectra)          & 9   & 4, 5\\
UGC~7069           & 13.4       & $0.08\pm{}0.014$ (spectra)       & --  & 6\\
Antennae           &  7.1       & 0.04    (X-ray)    & 8   & 7, 8, 9\\
NGC~4485/4490       & 1.0        & $<0.4$   (SDSS spectra)      & 8   & 7, 10, 11\\
NGC~3395/3396       & --    & $0.07^{+0.03}_{-0.01}$ (NGC~3395, X-ray), $0.05^{+0.04}_{-0.01}$ (NGC~3396, X-ray) & 7   & 12\\
The Mice (Arp~242) & 8.8        & 0.3   (X-ray)      & 5   & 13, 14\\
NGC~3256            & 44         & 1.0   (spectra of HII regions)      & 14  & 7, 15, 16\\
NGC~1313            & 1.4        & 0.1$-$0.2 (spectra)  & 2   & 17, 18, 19\\
NGC~4559            & --        & $0.05-0.2$ (Geneva STs), $0.2-0.4$ (Padua STs), $0.3^{+0.3}_{-0.2}$ (X-ray)  & 2   & 20\\
Holmberg~II        & 0.07$-$0.1 & 0.1   (spectra)      & 1   & 21, 22\\
\noalign{\vspace{0.1cm}}
\hline
\end{tabular}
\end{center}
\footnotesize{ $^{\rm a}$ The metallicities collected in this Table come from different measurement methods. For each entry we indicate the type of measurement  within parentheses. See the references for details. $^{\rm b}$ 1. Mayya et al. (2005); 2. Fosbury \& Hawarden (1977); 3. Wolter \& Trinchieri (2004); 4. Higdon \& Wallin (1997); 5. Giordano et al. in preparation; 6. Ghosh \&{} Mapelli (2008); 7. Grimm, Gilfanov \&{} Sunyaev (2003) and references therein; 8. Fabbiano et al. (2004); 9. Fabbiano, Zezas \& Murray (2001); 10. Pilyugin \&{} Thuan (2007); 11. Fridriksson et al. (2008); 12. Brassington, Read \&{} Ponman (2005); 13. Hunter et al. (1986); 14. Read (2003); 15. L\'ipari et al. (2000); 16. Lira et al. (2002); 17. Ryder \& Dopita (1994); 18. Ryder (1993); 19. Colbert et al. (1995); 20. Soria et al. (2005) and references therein; 21. Walter et al. (2007); 22. Dewangan et al. (2004).
}
\end{table*}
%%%%%%%%%%%%%%%%%%%%%%%%%%%%%%%%%%%%%%%%%%%%%%%%%%%%%%%%%%%%%%%%%%%%%%%%%%%%%
The model presented in this Letter works quite well for the Cartwheel (Section~3). Is it possible to check it for other galaxies? In Table~2 we have listed a small sample of galaxies which have interesting properties for this study. This sample includes galaxies which host at least one ULX and for which metallicity measurements are available (apart from the case of UGC~7069, for which X-ray measurements are currently unavailable, but which is quite similar to the Cartwheel for many aspects). 
In particular, the sample includes three of the best studied ring galaxies (Cartwheel, AM~0644-741 and UGC~7069), seven galaxies with a large number of ULXs ($\gtrsim{}2$ ULXs with  $L_{\rm X}\gtrsim{}10^{39}$ erg s$^{-1}$) and the dwarf irregular galaxy Holmberg II (HoII), which hosts a single, very bright ULX ($L_{\rm X}\gtrsim{}10^{40}$ erg s$^{-1}$, Dewangan et al. 2004). 
%We do not pretend  this sample to be complete: it has been selected on the basis of the existence of at least one ULX per galaxy and of the availability of metallicity measurements (apart from the case of UGC~7069, for which X-ray measurements are currently unavailable, but which is quite similar to the Cartwheel for many other aspects). 
%We do not pretend  this sample to be complete: a more complete one will be considered in a forthcoming paper.
%%%%%%%%%%%%%%%%%%%%%%%%%%%%%%%%%%% FIGURE 1 %%%%%%%%%%%%%%%%%%%%%%%%%%%%%%%%%%
%\begin{figure}
%\center{{
%\epsfig{figure=ULX_Z.eps,height=7cm}
%\epsfig{figure=ULX_SF.eps,height=7cm}
%}}
%\caption{\label{fig:fig1}
%Top (bottom) panel: number of ULXs versus metallicity (SFR) for the galaxies listed in Table~2.
%}
%\end{figure}
%%%%%%%%%%%%%%%%%%%%%%%%%%%%%%%%%%%%%%%%%%%%%%%%%%%%%%%%%%%%%%%%%%%%%%%%%%%%%%%
%%%%%%%%%%%%%%%%%%%%%%%%%%%%%%%%%%% FIGURE 2 %%%%%%%%%%%%%%%%%%%%%%%%%%%%%%%%%%
\begin{figure}
\center{{
\epsfig{figure=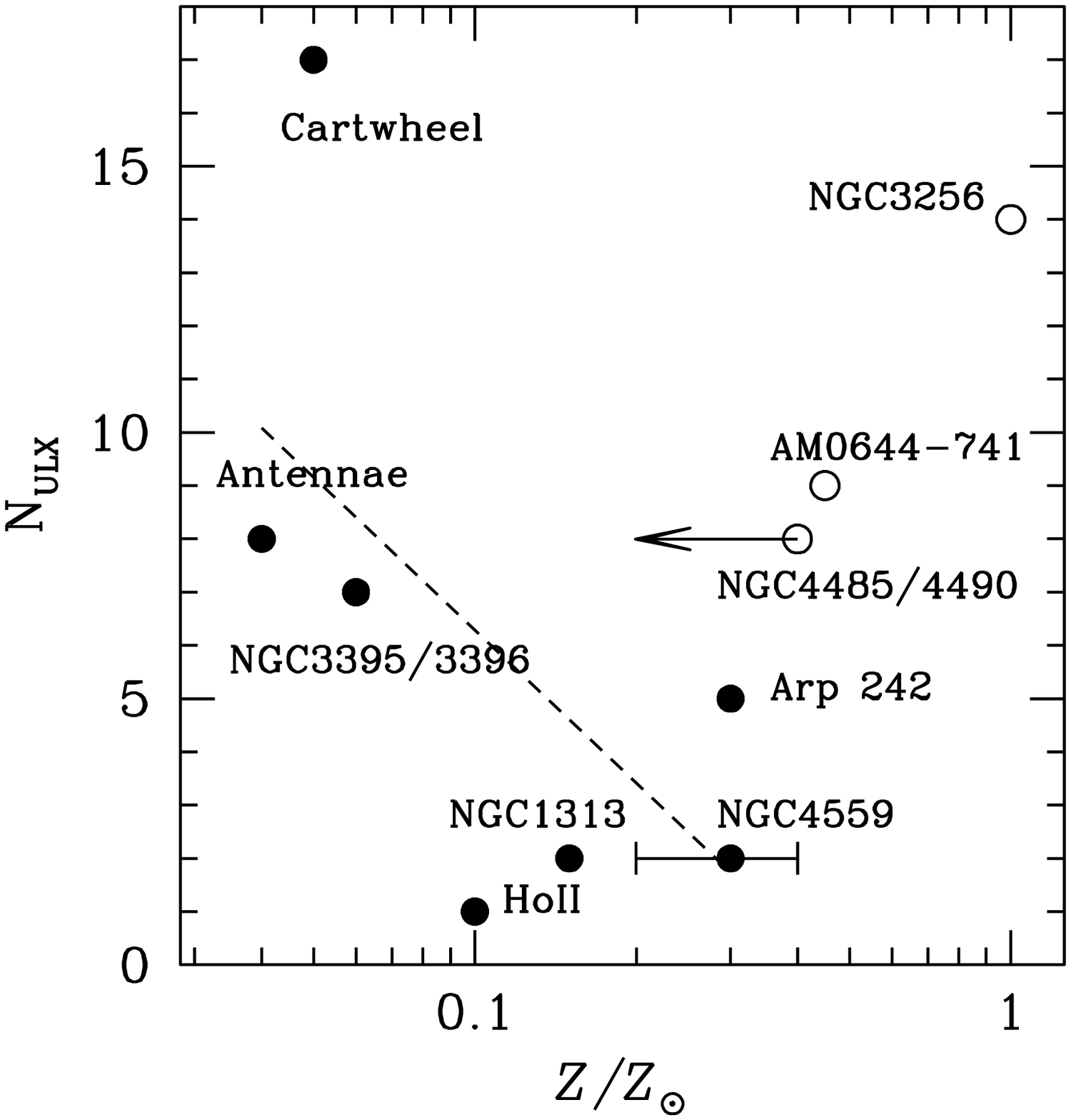,height=6cm}
\epsfig{figure=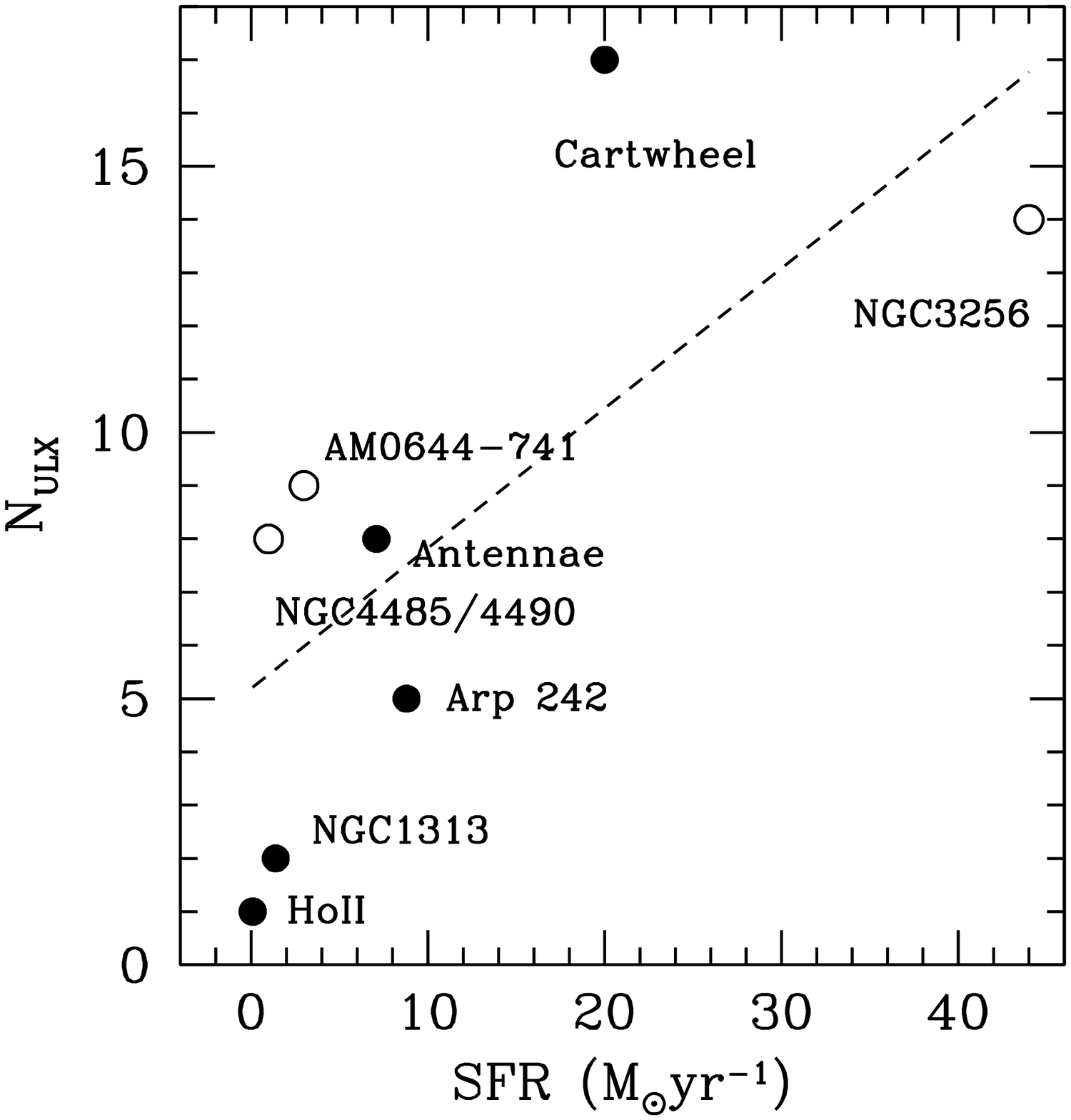,height=6cm}
}}
\caption{\label{fig:fig1}
Top (bottom) panel: number of ULXs versus metallicity (SFR) for the galaxies listed in Table~2. The open circles correspond to NGC~3256, AM~0644-741 and NGC~4485/4490 (see Section~4), the filled circles to all the other galaxies listed in Table~2. Dashed lines: linear fits.
}
\end{figure}
%%%%%%%%%%%%%%%%%%%%%%%%%%%%%%%%%%%%%%%%%%%%%%%%%%%%%%%%%%%%%%%%%%%%%%%%%%%%%%%

Fig.~\ref{fig:fig1} shows N$_{\rm ULX}$ versus the metallicity $Z$ (top panel) and the SFR (bottom panel) for the galaxies of our sample (filled and open circles). 
The galaxies indicated with open circles in Fig.~\ref{fig:fig1} do not have
metallicity measurements suitable for this study. In fact, the value of the metallicity needed in our model is the one the galaxy had before the beginning of the SF burst which produced the massive BHs.
%In Fig.~\ref{fig:fig1} 
%the galaxies listed in Table~2 are represented by filled circles, except for
%NGC~3256, AM~0644-741 and NGC~4485/4490, which are indicated by open circles. This different notation is due to the fact that the metallicity measurements of the galaxies indicated by open circles are not suitable for this study\footnote{We remind that the value of the metallicity we need for our model is the one the galaxy had before the beginning of the SF burst which produced the massive BHs.}. 
In the case of AM~0644-741 the metallicity measurement comes from the bulge (Giordano et al. in preparation), which is dominated by old stars, whereas there are no measurements for the star forming outer ring. In the case of NGC~4485/4490, the value of $Z$ quoted in Table~2 is likely an upper limit, as the method used by Pilyugin \&{} Thuan (2007) does not work for metallicities $Z\lesssim{}0.4$. Finally, NGC~3256 has already undergone a strong burst of SF and it is difficult to find HII regions which are still unpolluted in this galaxy (L\'ipari et al. 2000), whereas in ring galaxies or in interacting galaxies it is relatively easy to find regions where the SF just started. 
The filled circle for NGC~4559, corresponding to $Z=0.3\,{}Z_\odot{}$, comes from X-ray measurements (Cropper et al. 2004), and the error bar comes from the estimate ($0.2\lesssim{}Z/Z_\odot{}\lesssim{}0.4$, Soria et al. 2005) obtained with the Padua stellar tracks (STs). However, when the Geneva STs are used instead of the Padua STs, the estimated metallicity of NGC~4559 is quite lower ($0.05\lesssim{}Z/Z_\odot{}\lesssim{}0.2$, Soria et al. 2005).
%The huge error bars for the metallicity of NGC~4559 (which must not be confused with NGC~1313) come from the uncertainties in the stellar track (ST) methods: the Geneva ST and the Padua ST give $0.05\lesssim{}Z/Z_\odot{}\lesssim{}0.2$ and $0.2\lesssim{}Z/Z_\odot{}\lesssim{}0.4$, respectively (Soria et al. 2005). The filled circle for NGC~4559, corresponding to $Z=0.3\,{}Z_\odot{}$, comes from X-ray measurements (Soria et al. 2005). 
The absence of error bars for most of metallicities in the top panel of Fig.~\ref{fig:fig1} means only that the measurements are highly uncertain and/or that there are no estimates of the error (see Table~2 and references therein for details).
In general, the metallicity measurements reported in Table~2 are uncertain and/or obtained with very different methods, from spectral analysis to X-ray data. Thus, they are quite difficult to compare with each other and the top panel of Fig.~\ref{fig:fig1} must be considered {\it cum grano salis}. New, homogeneous metallicity  measurements are required, in order to test a possible relation between N$_{\rm ULX}$ and $Z$. 

Taking into account all these strong {\it caveats}, the top panel of Fig.~\ref{fig:fig1} suggests a correlation between N$_{\rm ULX}$ and $Z$. If we exclude AM~0644-741, NGC~3256 and NGC~4485/4490 for the reasons mentioned above, we find a correlation:
\begin{equation}\label{eq:corr}
N_{\rm ULX}=\beta{}\,{}\log{}_{10}(Z/Z_\odot{})+\gamma{},
\end{equation}
where $\beta{}=-9.53$ and $\gamma{}=-3.25$. The correlation is shown by the dashed line in the top panel of Fig.~\ref{fig:fig1}.
%where $\beta{}=-9.346$ and $\gamma{}=-2.923$. The correlation is shown by the dashed line in the top panel of Fig.~\ref{fig:fig1}.
The bottom panel of Fig.~\ref{fig:fig1} shows that there is an evident correlation between SFR and N$_{\rm ULX}$ for our sample, in agreement with previous studies (Grimm, Gilfanov \&{} Sunyaev 2003; Gilfanov, Grimm \&{} Sunyaev 2004a, 2004b, and references therein). Excluding NGC~3395/3396 and NGC~4559, for which there are no estimates of the global SFR, we obtain the following linear relation.
%For our sample (including AM~0644-741, NGC~3256 and NGC~4485/4490, but excluding NGC~3395/3396  and NGC~4559 which do not have any estimate of the global SFR) we obtain the following linear relation.
\begin{equation}\label{eq:corr2}
N_{\rm ULX}=\delta{}\,{}\,{}\textrm{SFR}\left[M_\odot{}\,{}{\rm yr}^{-1}\right]+\zeta{},
\end{equation}
where $\delta{}=0.26$ and $\zeta{}=5.19$. The correlation is shown with a dashed line in the bottom panel of Fig.~\ref{fig:fig1}.
%where $\delta{}=0.2571$ and $\zeta{}=5.38$. The correlation is shown by the dashed line in the bottom panel of Fig.~\ref{fig:fig1}.

In conclusion, our sample confirms the existence of a correlation between SFR and N$_{\rm ULX}$ and suggests a possible anti-correlation between $Z$ and N$_{\rm ULX}$. The fact that galaxies which host a large number of ULXs have often low metallicity indirectly supports the scenario proposed in Sections 2 and 3. In particular, three of the considered galaxies (i.e. the Cartwheel, the Antennae and NGC~3395/3396) host a large number of ULXs (N$_{\rm ULX}\ge{}7$) and have very low metallicity ($Z\sim{}0.06\,{}Z_\odot{}$). Thus, we can apply also to the Antennae and to NGC~3395/3396 the method\footnote{At present we cannot apply our method to galaxies with $Z\gtrsim{}0.06\,{}Z_\odot{}$, because we do not know which is the minimum initial mass for which stars with  metallicity $0.5\gtrsim{}Z/Z_\odot{}\gtrsim{}0.06$ end their life by directly collapsing into BHs. However, it is likely that massive BHs are formed also by stars with metallicity slightly higher than $0.06\,{}Z_\odot{}$. Further studies are needed, to constrain the upper limit of stellar metallicity for which massive BHs can form.}  used for the Cartwheel in Section~3. For the Antennae, adopting a SFR of 7.1 $M_\odot{}\,{}$yr$^{-1}$, N$_{\rm ULX}=8$ (Table~2) and $t_{\rm burst}=10^7$ yr, we obtain $\epsilon{}_{\rm BH}=1.8\times{}10^{-4}$ and $\epsilon{}_{\rm BH}=9\times{}10^{-5}$ for the Salpeter and the Kroupa IMF, respectively. In the case of  NGC~3395/3396, the SFR is uncertain. Adopting a SFR of 6 $M_\odot{}\,{}$yr$^{-1}$ (derived from the correlation between SFR and X-ray luminosity in eq. 14 of Ranalli, Comastri \&{} Setti 2003),  N$_{\rm ULX}=7$ (Table~2)  and  $t_{\rm burst}=10^7$ yr, we obtain $\epsilon{}_{\rm BH}=1.9\times{}10^{-4}$ and $\epsilon{}_{\rm BH}=9\times{}10^{-5}$ for the Salpeter and the Kroupa IMF, respectively. These values are similar to those obtained for the Cartwheel and to the estimates of $\epsilon_{\rm exp}$ reported in Table~1. Thus, our method gives reasonable results not only for the Cartwheel but also for other interacting galaxies. 
%%%Adopting a reasonable SFR of 10 $M_\odot{}\,{}$yr$^{-1}$,  N$_{\rm ULX}=7$ (Table~2)  and  $t_{\rm burst}=10^7$ yr, we obtain $\epsilon{}_{\rm BH}=1.1\times{}10^{-4}$ and $\epsilon{}_{\rm BH}=6\times{}10^{-5}$ for the Salpeter and the Kroupa IMF, respectively. These values are similar to those obtained for the Cartwheel and to the estimates of $\epsilon_{\rm exp}$ reported in Table~1. Thus, our method gives reasonable results not only for the Cartwheel but also for other interacting galaxies. 

We stress that the fact that many galaxies which host ULXs (e.g. NGC~4485/4490, Arp~242, NGC~1313, Holmberg II, etc.) have metallicity (slightly) higher than that of the Cartwheel  does not contradict our scenario for massive BH formation. In fact, the metallicity we measure now is not necessarily the same as the metallicity of the parent molecular cloud, where the massive BHs formed. It is likely that the metallicity we measure now has been partially polluted by the episode of SF which generated the massive BHs. Only for the collisional ring galaxies, where the dynamical evolution of the ring is strongly coupled to the SF history, it is relatively easy to measure the metallicity of the pre-starburst gas (i.e. the gas which resides in the outer ring).

\section{Conclusions}
%In this Letter 
%We have proposed
%We have investigated 
Low-metallicity ($Z\lesssim{}0.05\,{}Z_\odot{}$) massive ($\gtrsim{}40\,{}M_\odot{}$) stars might end their life by directly collapsing into massive BHs ($30\lesssim{}m_{\rm BH}/M_\odot{}\lesssim{}80$, M92; H03). Such massive BHs might power most of the observed ULXs in low-metallicity galaxies (such as the Cartwheel and the Antennae). 
%Furthermore, the expected number of such massive BHs in low-metallicity galaxies (such as the Cartwheel and the Antennae) is consistent with the observed number of ULXs and with their expected efficiency. 
In support of this scenario, the data listed in Table~2 suggest an anti-correlation between the number of ULXs and the metallicity of the host galaxy. 

However, many open questions and uncertainties remain. First of all, the final stellar masses reported by M92 are still debated: a similar study by Portinari, Chiosi \&{} Bressan (1998) finds  sensibly lower masses. Belczynski, Sadowski \&{} Rasio (2004) also find lower initial masses for the BHs ($\sim{}20-30\,{}M_\odot{}$), but investigate the possibility of increasing the mass of the BH (up to $\sim{}80\,{}M_\odot{}$) via binary mergers.
Furthermore, the models considered in M92 and in Fryer (1999) neglect some important effect, such as the rotation and the possible binarity of the progenitor. Accounting for the  binarity of the progenitor likely induces a factor of 2 uncertainty in our results. In addition, the model considered here does not include the possibility of pair instability supernovae (PISs). PISs probably do not play a role for stars with $Z\sim{}0.05\,{}Z_\odot$ (H03). On the other hand, even assuming (as a strong upper limit) that all stars with mass $\ge{}100\,{}M_\odot{}$ do not leave any remnant, due to a PIS, our estimates of $\epsilon{}_{\rm BH}$ change by less than 10 per cent.
% binarity: factor 0.5.
%pair instability supernovae.
%may not account the brighter ULXs

The lack, paucity or uncertainty
in the metallicity measurements make hard to test our model.
% (see Section 4).
%At the moment, one of the biggest problems of this method is that metallicity data are often uncertain or completely unavailable. This fact makes hard to test our model (see Section 4). 
Moreover, the metallicity needed in our model is that of the molecular clouds before the pollution from the first generation of supernovae, as very massive stars ($>40\,{}M_\odot{}$) collapse into BHs before the explosion of the first supernovae in the parent cluster. Thus, the metallicity measured today is likely higher than the value we should consider in our model. Only for some types of galaxies, such as the ring galaxies, where the SF history has a clear connection with the geometry of the system, it is possible to measure a pre-starburst value of $Z$, suitable for our purposes.

%%%We also note that the analytical estimates reported here may be affected by variations of metallicity across galaxies and multiple star formation episodes.

%the metallicity which matters is always before the main pollution from supernovae, as very massive stars ($>40\,{}M_\odot{}$) collapse into BHs before the explosion of the first supernovae in the parent cluster. Thus, the metallicity measured today represent likely an upper limit.

\section*{Acknowledgments}
We thank E.~Ripamonti, R.~Decarli, A.~Wolter, P.~Englmaier, L.~Giordano and A.~Bressan for useful discussions and we acknowledge the anonymous referee for his helpful comments. 
MM acknowledges support from the Swiss
National Science Foundation, project number 200020-109581/1.
%(Computational Cosmology \&{} Astrophysics). 
MC and LZ acknowledge financial support from INAF through grant PRIN-2007-26.

\end{document}